\documentclass[12pt,a4paper]{article}
\usepackage{latexsym,amsfonts,amsmath,amssymb}
\usepackage{ifthen}
\newtheorem{theorem}{Theorem}[section]

\newtheorem{proposition}{Proposition}[section]

\begin{document}
%\title{There is no generalization of Wooters \& Fields formula for mutually unbiased bases}%
\title{There is no generalization of known formulas for mutually unbiased bases}%

\author{Claude Archer\\
%\address{
Universit\'e Libre de Bruxelles,\\
C.P.165/11 -Physique et Math\'ematique \\
Facult\'e des Sciences Appliqu\'ees \\
avenue F.D. Roosevelt 50,\\
1050 Bruxelles, Belgium \\
\it{carcher@ulb.ac.be}}

%\address{Universit\'e Libre de Bruxelles}%
%\email{carcher@ulb.ac.be}%

%\thanks{}%
%\subjclass{}%
%\keywords{}%

\date{}%
%\dedicatory{}%
%\commby{}%
% ----------------------------------------------------------------
\maketitle
\begin{abstract} In a quantum system having a finite number $N$ of orthogonal states,
two orthonormal bases $\{a_i\}$ and $\{b_j\}$  are called mutually unbiased if all inner products $\langle a_i|b_j \rangle $
 have the same modulus  $1/\sqrt{N}$.
This concept appears in several quantum information problems.
  The number of pairwise mutually unbiased bases
 is at most $N+1$  and various constructions of $N+1$ such bases have been found when $N$ is a power
 of a prime number. We study families of formulas that generalize these constructions to arbitrary dimensions
 using finite rings.
We then prove that there exists a set of $N+1$ mutually unbiased bases described by such formulas, if and
only if $N$ is a  power of a prime number.

%In 1988, Wooters \& Fields have found a formula that
%provides a complete set of $d+1$ mutually unbiased bases when $d$ is an odd prime-power dimension.
%This formula uses the finite field of order $d$. We study famillies of formulas that generalize this on
%finite rings for any dimension $N$.
% We then prove that if $d$ is not a prime power, such formulas will
% not provide a complete set of $d+1$ mutually unbiased bases %%but at most  1+min(p_1^{e_1},\,i=1\ldots r)$ mutually unbiased bases.
%We then prove that for a dimension d, there exists a complete set of mutually unbiased bases
%described such formula if and only if d is a prime power.
\end{abstract}
\newpage
% ----------------------------------------------------------------
\section{Introduction}
\subsection{Definitions and previous results}
%Let $\{a_i\}$ and  $\{b_i\}$ be For $i,j \in \{1\ldots N\}$

In the N-dimensional Hilbert space $\mathbb{C}^{N}$, two orthonormal bases $\{a_i\}_{1\leq i\leq N}$ and
$\{b_j\}_{1\leq j\leq N}$  are called \emph{mutually unbiased} if all inner products $\langle a_i|b_j \rangle $
 have the same modulus  $|\langle a_i|b_j \rangle |=1/\sqrt{N}$. A set of mutually unbiased bases % (\emph{\textbf{$MUB$}} for short)
 is a set of  orthonormal bases  which are pairwise mutually unbiased.
%% We often abbreviate \emph{mutually unbiased bases} as MUB.
 In various physical situations (see subsection \ref{phys}), the problem is to find the maximal number
of mutually unbiased bases.
The following result is due to W.K. Wootters and  B.D. Fields but it has been obtained  independently by
A. R. Calderbank, P. J. Cameron, W. M. Kantor and J. J. Seidel.
%Determine the maximal number K of orthonormal bases in a D-dimensional Hilbert space, which are
%mutually unbiased in the following sense: If eik denotes the ith vector of the kth basis, all scalar products
%$\langle  eik,ejn \rangle $ with $k\neq n$ have the same absolute value (namely D-1/2).
%It is known that if D is the power of a prime, K=D+1 can be reached, but this is not known for any
%other composite number. So the problem is already to decide whether there exist K=7 mutually unbiased
%bases in D=6 dimensions.

\begin{theorem} \label{WoFi}(\cite{WF}, \cite{CalderCam})
\begin{itemize}
\item In dimension $N\geq 2$, the number of mutually unbiased bases is at most $N+1$.
\item If $N$ is a power of a prime number then there exist $N+1$ mutually unbiased bases.
\end{itemize}
\end{theorem}
In dimension $N$, a set of mutually unbiased bases
  is called \emph{complete} if it contains $N+1$ bases.
If $N$ is not a prime power, it is not known whether such a complete set exists, even for $N=6$.
Originally, constructions of $N+1$ mutually unbiased bases in dimension $N$, were based on the arithmetic of a  field, where addition and multiplication are invertible (\cite{WF}). There exists a field with $N$
elements if and only if $N$ is a power of a prime number. Nevertheless, new constructions
%of $N+1$ mutually unbiased bases
have been recently obtained (see \cite{RotKlap}) using the arithmetic of rings, where multiplication is not invertible. Since there exist rings of $N$ elements for any $N$, is it possible to use finite rings to  construct $N+1$ mutually unbiased bases for arbitrary dimensions ? We will address this issue here.
First we will generalize known constructions from \cite{WF} and \cite{RotKlap} to any finite ring. Then
we will prove that for dimensions $N$ that are not prime powers, there does not exist a complete set of
mutually unbiased bases described  by this generalization.

%%%%%%%%%%%%%%%%%%%%%%%%%%%%%%%%%%%
%%\subsection{Physical origins of the problem}\label{phys}
\subsection{Applications to Quantum Information}\label{phys}
% Je suggère un titre tel que "Applications to quantum information"
%ou plus simplement, et ça me plaît davantage: %"MUB and Quantum information".
Mutually unbiased bases (\emph{\textbf{$MUB$}} for short)  have recently been considered with
an increasing interest because of the central role they play in specific quantum information tasks.
MUB are related,  among others,  to state estimation and to protocols of quantum cryptography.

\emph{State estimation.}
Mutually unbiased bases play an important role in state estimation of (relatively) large
ensembles of identical prepared quantum systems. 
MUB allow us to minimize the number of measurements needed to estimate a quantum state.
The density matrix of an $N$-dimensional quantum state is determined by $N^2-1$ real
parameters. Hence, at least $N+1$ measurements are needed to
re-construct such a density matrix. One can show that  $N+1$
measurements are \emph{sufficient} if these measurements are MUB
(\cite{Ivanovic}). The reason is that if two measurement bases $B_1$ and
$B_2$ are mutually unbiased, then the information revealed by the
outcomes of these measurements are independent. Other optimality properties of MUB
with respect to state estimation are described in \cite{WF}. \\

%\emph{State determination.} If one wants to minimize the number of measurements needed to estimate a
%quantum state, an optimal way to proceed is to choose measurements associated to mutually unbiased bases.
%%The concept of mutually unbiased bases appeared in the problem of minimizing the number of
%%measurements required for the estimation of a quantum state.
%The density matrix of an ensemble of N-state quantum systems is determined by $N^{2}-1$ real parameters.
%The statistics of a measurement of these states yields $N-1$ real parameter. Hence at least $N+1$
%measurements are needed to determine the density matrix.
%Ivanovic (see \cite{Ivanovic}) has shown that the density matrix can be determined by the statistics of exactly
%$N+1$ measurements if these correspond to mutually unbiased bases.
%If two bases $A$ and $B$ are mutually  unbiased then the measurement corresponding to basis
%$A$ do not duplicate information contained in  measurement associated to $B$.
%Other properties of mutually unbiased bases with repect to state estimation are described in \cite{WF}.\\

 \emph{Quantum cryptography.}
The protocol BB84 of Bennett and Brassard (\cite{BB84}) for quantum key distribution, used
with the one-time pad encryption,
is the first
cryptographic protocol whose security does not depend on the assumption that an eavesdropper has
a limited computational power. Its security is guaranteed by  Heisenberg's  uncertainty principle.
%%the laws of Quantum Mechanics

%By Heisenberg's  uncertainty principle , if an eavesdropper Eve observes a message coded into a quantum
%system, she modifies the system itself and this can be noticed by Alice and Bob.

MUB are the basic algebraic structure underlying  $d$-dimensional analogues
of  the BB84 protocol (\cite{Cerf}). It is  precisely the use of such
 bases which allows these  protocols to make the intervention of a potential eavesdropper detectable.
%The protocol BB84 uses mutually unbiased bases
 %to send a secret key over a public channel in a secure way.
Alice and Bob agree on a set of $t$ orthonormal bases and
Alice sends to Bob a state $a_i$ prepared in a basis $A=\{a_i\}$ taken among the $t$ bases.
Bob chooses a basis in one of the $t$ bases to measure this state. If Bob chooses the right basis,
 he finds the good value.  Now suppose an eavesdropper Eve has chosen $E=\{b_j\}$ to measure the state.
 Eve obtains $b_j$ as result  with probability $|\langle a_i|b_j \rangle |^{2}$. Since for security purpose,
 one wants that when $E$ is the wrong base, Eve  gets no
 information on the original state, we should require that all probabilities $|\langle a_i|b_j \rangle |^{2}$ are equal
 and hence equal to $1/N$ ( i.e. $A$ and $E$ are mutually unbiased). Moreover, as the probability
 to choose a wrong basis is $1/t$, one wants to take the largest possible number of mutually unbiased bases.
 It is also known that a protocol using a larger number of mutually unbiased bases can tolerate
a higher error level in the channel (see \cite{Cerf}).\\

%Mutually unbiased bases also appear in the Mean King problem.
\emph{Wigner functions} Pure or mixed quantum states are usually represented by the density matrix. However,
there is an alternative description in terms of the Wigner function. Several authors have proposed to
define a Wigner function for discrete systems having $N$ degrees of freedom. It appears that the discrete Wigner function defined in \cite{wigner} requires the existence of $N+1$ mutually unbiased bases.\\

Finally, MUB have also been shown to be relevant to the mean
king's problem,  see \cite{MeanKingAE} and references therein.
An interesting source for recent results and for references is the problem page in Quantum
Information at TU Braunschweig, located at \emph{http://www.imaph.tu-bs.de/qi/problems}.

\section{Formulas for mutually unbiased bases}\label{formulas}

A unitary transformation maps a set of MUB to a set of MUB. Hence, it is not restrictive to consider only sets
 $X$ of MUB containing the standard basis $\{e_k\}_{1\leq k\leq N}$ since it is always possible to choose a unitary transformation $U$ that maps a given orthonormal basis in $X$ to $\{e_k\}$ so that $U(X)$ is a set of MUB containing $\{e_k\}$.
%Observe also that if a basis $r$ is unbiased with respect to the standard basis then the coordinates of its %vectors must be expressed as $(v_k ^{(r)})_l=\frac{1}{\sqrt{N}}e^{i \Theta (r,k,l)}$ where $\Theta(r,k,l) $ %belongs to $[0,2\pi ]$.
If a basis $\{v_k\}$ is unbiased with respect to the standard basis $\{e_k\}$
(i.e. $|\langle e_i|v_j \rangle |=1/\sqrt{N}$) then $|(v_k )_l|=|\langle e_l|v_k \rangle |=
    1/\sqrt{N}$. Hence the coordinates of its vectors must be expressed as
    $(v_k )_l=( e^{i \Theta (k,l)})/\sqrt{N}$ where $\Theta(k,l) $ belongs to $[0,2\pi ]$.\\
For $N=p^n$ where $p$ is a prime number and $n$ a positive integer, there always exist $N+1$ mutually
unbiased bases. We describe here the constructions from  \cite{WF} and \cite{RotKlap} for these dimensions.

\subsection{Odd prime powers dimensions}
%see \cite{WF} and \cite{Ivanovic}.
%\subsection{Wooters \& Field formulas}

Let the superscript $r$ denotes the basis, $k$ the vector in the basis and $l$ the component.
The standard basis is $(v_k ^{(0)})_l=\delta _{kl}$ for $k,l=0,1,\ldots ,N-1$. If $N=p^n$ for a prime
number $p\neq 2$, the other $N=p^n$ such bases given in \cite{WF} are,
%\begin{equation}
%(v_k ^{(0)})_l=\delta _{kl}\qquad k,l=0,1,\ldots ,N-1.
%\end{equation}

%%\begin{equation}
%%(v_k ^{(r)})_l=\frac{1}{\sqrt{N}}e^{(2\pi i/N)(rl^2 + kl)}\qquad k,l=0,1,\ldots ,N-1, \quad r=1,2,\ldots ,N
%%\end{equation}
%%For $N=p^n$, the other $N$ bases are,
\begin{equation}\label{pnodd}
(v_k ^{(r)})_l=\frac{1}{\sqrt{N}}e^{(2\pi i/p)Tr(rl^2 + kl)}\qquad r,k,l \in \mathbb{F}_{p^n}%%,\quad r\neq 0
\end{equation}
where $\mathbb{F}_{p^n}$ is the finite field with $p^n$ elements and
where $Tr$ denotes the trace map from $\mathbb{F}_{p^n}$ into the prime field $\mathbb{F}_{p}$.
For $p\geq 5$ odd, a new formula has been proposed in \cite{RotKlap} where the polynomial $rl^2 + kl$ is
replaced by $(l+r)^3 +k(l+r)$.
The trace map is a linear map from  $\mathbb{F}_{p^n}$, regarded as a vector space,
into $\mathbb{F}_{p}$. In the language of group theory, linear maps are group homomorphisms ( i.e. maps
that preserve sums) . The trace map induces
a homomorphism from the additive group of $\mathbb{F}_{p^n}$ into the multiplicative group
$\mathbb{C}^*$ of complex numbers, defined by $x\to e^{(2\pi i/p)Tr(x)}$. \\

%Since the trace map is linear, it induces
%a homomorphism from the additive group of $\mathbb{F}_{p^n}$ into the multiplicative group
%$\mathbb{C}^*$ of complex numbers, defined by $x\to e^{(2\pi i/p)x}$.\\
%%the notation $r\neq 0$ is ambiguous since $r$ takes as value the zero element of the field $\mathbb{F}_{p^n}$
%%(the corresponding basis is calle d the Fourier transform base).

%\beq \label{ivanovic}
% $g.\overline{u}=\overline{u}.g^{\xi (u)}$
% \eeq
%\beq \label{rule2}
% \overline{u}.\overline{v}=\overline{uv} [u,v]
% \eeq
\subsection{Even prime powers dimensions}
%%\subsection{A construction with finite rings}
For $N=2^n$, W.K.Wootters and B.D.Fields (\cite{WF}) have used an \emph{ad hoc} construction that may be reformulated
in a finite ring $R$ whose $4^n$ elements are sequences $(x_1,\ldots,x_n)$ with $x_i\in \mathbb{Z}_4$.
A much easier construction has been found recently by A. Klappenecker and M. Roetteler
(\cite{RotKlap}) using the Galois ring $R=GR(4,n)$. Let $Tr$ denotes the trace map from
$GR(4,n)$ into $\mathbb{Z}_4$. Once again $T:x\to e^{(2\pi i/4)Tr(x)}$ is a group homomorphism from
$GR(4,n),+$ into $\mathbb{C}^*$. The $2^n$ indexes are the elements of
$\mathit{T_n}$, the  $Teichm \ddot{u} ller$ set of $GR(4,n)$ and the $2^n$ bases described by
\begin{equation}\label{galois}
(v_k ^{(r)})_l=\frac{1}{\sqrt{2^n}}e^{(2\pi i/4)Tr((r+2k)l) }\qquad r,k,l \in \mathit{T_n}\subset GR(4,n)
\end{equation}
%%where $\mathit{T_n}$ is the Teichmüller set and
together with the standard basis , form a complete set of mutually unbiased bases
of $\mathbb{C}^{2^n}$ (see \cite{RotKlap}).

%The standard basis together with the $2^n$ bases of equation \ref{galois} form a complete set of mutually
%unbiased bases of $\mathbb{C}^{2^n}$ (see \cite{RotKlap}).

\subsection{How to generalize these formulas ?}
%%\subsection{How to generalize known formulas to finite rings ?}
Formulas (\ref{pnodd}), (\ref{galois}) as well as others in \cite{RotKlap}, share many common characteristics.
First of all, the indexes $l,k,r$ respectively for components, vectors and bases are taken
in a finite ring $R$.  Both formulas link the indexes in $R$ to complex coordinates
by a function $f:(r,k,l)\to T(P(r,k,l))$ where $P$ is a polynomial and $T$ is a homomorphism from $R,+$
into $\mathbb{C}^*$. We will generalize these characteristics as follows :

\begin{enumerate}
%%%\item Finite field $\to$ Finite ring.
\item \emph{The functions $f:(r,k,l)\to T(P(r,k,l))$}.
 We consider a much larger class of functions that
we call \emph{functions preserving a direct sum decomposition of $R$} (see section \ref{fpres}).
%From a fixed Trace map, to various group homomorphisms.
%To create a link In formulas \ref{pnodd} and \ref{galois} the trace map is a linear map from  $\mathbb{F}_{p^n}$, regarded as a vector space,
%into the multiplicative group of complex number. In the language of group theory, "linear maps" ( i.e maps
%that preserve sums) are called group homomorphisms. The trace map in formula \ref{galois} is a group
%homomorphism from the group
%viewedthe vector space A group homomorphism $T$ from $R,+$ into a group $G,\cdot$ maps
%$S_{p_i}$ into since for ou purpose
%Constant Trace map $\to$ $t$ distinct (one for each base)
%linear map $T^{t}$ from  $R,+$ into $\mathbb{C}_0, \cdot$.
%\item a fixed polynomial $P$ in the index $t,l,k$ $\to$ different polynomials $P^{(t)}$ in the index $l,k$.
% whose coefficients depend on $t$ ( a different polynomial $P^{(t)}$ for each basis ).

\item \emph{The set $S$ of indexes.}
%ù$S$ such that $|I|$ divides $|A|$ $\to$ set of indexes
%%$S=\prod_i S_i$ such that $|S_i|$ divides $|A_i|$.
For formula (\ref{pnodd}) the set of indexes is the whole $R$ while for formula (\ref{galois}),
it is a remarkable subset of $R$. %% of size $\frac{|R|}{2^n}$. 
We will see in subsection \ref{Teich} that these subsets may be defined for every ring $R$ as sets
closed under multiplication and transversal to a nilpotent ideal of $R$.

\item \emph{Distinguish the index $r$}.
In formula (\ref{pnodd}) and (\ref{galois}), the non standard bases are indexed by $r$ that takes all possible values of a set of size $N$. This can only be done for dimensions $N$ for which there exist
 $N+1$ mutually unbiased bases. However there is, up to now, no result showing that this is
true if $N$ is not a prime power. Therefore, we propose to give up formulas that are
a uniform with respect to $r$ and to consider that each basis $r$ may be described by a different formula.
This means that for each $r$, we choose a different function $f_{r}:(k,l)\to f_{r}(k,l)$ into $\mathbb{C}^*$
such that the vectors in basis
 $r$ are described by
\begin{equation}\label{generalize}
 (v_k ^{(r)} )_l=f_{r}(k,l)\qquad k,l\,\in S
 \end{equation}
where each $f_r$ preserve a given decomposition $R_1\oplus R_2$ and $S\subset $ is as described in previous paragraph. For the case of polynomials $P$ and homomorphisms $T$, it  amounts to choose for each $r$, different polynomials $P_{r}$ and homomorphism $T_{r}$.

\end{enumerate}

\subsection{Properties of rings}\label{finite rings}
In this section we recall various properties of rings that are needed for this paper.\\

\emph{Direct sums of rings.}
Let $R_{+,\cdot}$ be a ring where addition is commutative but where multiplication is not necessarily
 commutative. If the additive group $R_{+}$ is the direct sum $R_1\oplus R_2$ of two subgroups then 
every element $r$ of $R$ can be written in a unique way as $r=r_1 +r_2$ where $r_i \in R_i$ and
 $R_1\cap R_2$ is reduced to the zero element. An element $r\in R$ may be represented as a couple
 $(r_1,r_2)$ and addition in $R$ corresponds to componentwise addition of couples. If moreover, $R_1$ and
 $R_2$ are two-sided ideals of $R$ (i.e. $r\cdot R_i=R_i=R_i \cdot r$ for every $r$ in $R$), then
 multiplication in $R$ is also reduced to componentwise multiplication of couples. Indeed, for $r_i \in R_i$,
 $r_1.r_2$ belongs to both $R_1$ and $R_2$ since these are two-sided ideals and thus
 $r_1.r_2=0=R_1\cap R_2$. For two ring elements $x$ and $y$,  we obtain for their product
 $x. y=(x_1 +x_2). (y_1 +y_2)= x_1.y_1 + x_1.y_2 + x_2.y_1 +x_2.y_2$ and since $x_i .y_j=0$ for $i\neq j$
 %. But for $i\neq j$, the two-sided ideal  property implies that $x_i .y_j$ belongs to $R_i\cap R_j$ which is zero. Hence, as annouced,
, we get $x. y=x_1.y_1 +x_2.y_2$.  Thus $(x.y)_i=x_i .y_i$.
Observe also that $a.r_1=(a_1 + a_2).x_1=a_1.x_1$ for every $a\in R$.
We say that the ring $R$ is the direct sums of its ideals $R_1$ and $R_2$.
These properties also hold when $R$ is the direct sum of more than two ideals.\\

%\emph{Polynomial functions in a ring.} A monomial on the set of variables $\{x_1,\ldots,x_k\}$ is a finite product
%of elements of this set. As for polynomials in a commutative ring, non commutative
%polynomials are defined as linear combinations of monomials but the ordering of factors in the monomial matters.
%Monomials $x_1.x_2$ and $x_2. x_1$ have to be considered as different as well as $r.
%different monomials and
% non commutative polynomial functions $R[x_1,\ldots,x_k]=\{\sum_j a_j m(x_1,\ldots,x_k)b_j \}$ where
%$m(x_1,\ldots,x_k)$ is a finite product of elements taken in the set $\{x_1,\ldots,x_k\}$ of non commuting
%variables.

\emph{Polynomial functions in a ring.}
If a ring is a direct sum, then let us show that polynomial functions in $R$ may be evaluated componentwise.
A monomial on the set of variables $\{x,y,\ldots\}$ is a finite product
of elements of this set.
In a commutative ring $R$, a polynomial $P(x,y,\ldots)+r_0$ is defined as a linear combination $P$ of
monomials (on $\{x,y,\ldots\}$  ) with coefficients in $R$ and of $r_0\in R$.
A polynomial $P+r_0$ defines a polynomial function
$(x,y,\ldots) \to P(x,y,\ldots)+r_0$ that maps n-uples of  $R^n$ to elements of $R$.
In the non commutative case, this definition is not very convenient since generally $x.r.y\neq r.x.y$ so that
products of polynomials would not in general be polynomials. Hence we prefer to define a non commutative
polynomial function as $P:(x,y,\ldots) \to r_0+\sum_k m_k(x,y,\ldots)$ where
$m_k(x,y,\ldots)=w_k(x,y,\ldots; a,b,\ldots)$ is a finite product of non commuting variables in
$\{x,y,\ldots\}$ and  of coefficients from a set $\{a,b,\ldots\}\subset  R$.

%polynomial function as $P:(x,y,\ldots) \to \sum_k w_k(x,y,\ldots; a,b,\ldots)$
 %where each $m_k$ is a finite product of non commuting variables in $\{x,y,\ldots\}$ and
% of coefficients from the set $\{a,b,\ldots\}\subset  R$.

Assume now that the ring $R$ is a direct sum $R=R_1\oplus R_2$ and for $r\in R$, let $r=r_1 +r_2$
be the corresponding decomposition. We have shown in previous paragraph that products in $R$ may
be performed componentwise. Thus, for each term $w_k$ of a polynomial function %we have
$w_k(x,y,\ldots; a,b,\ldots)$ is equal to $w_k(x_1,y_1,\ldots; a_1,b_1,\ldots)+
w_k(x_2,y_2,\ldots; a_2,b_2,\ldots).$
As $x_1.a_1=x_1.a$ and $a_1.x_1=a.x_1$ for $a,x\,\in R$ we may conclude that
 $w_k(x_i,y_i,\ldots; a_i,b_i,\ldots)$ is equal to $w_k(x_i,y_i,\ldots; a,b,\ldots)$, for $i=1,2$, so that
 $m_k(x,y,\ldots)=m_k(x_1,y_1,\ldots)+m_k(x_2,y_2,\ldots)$.
Hence, for a polynomial function $\bar{P}=P+r_0$ on $R_1\oplus R_2$, we have
$P(x,y,\ldots)= P(x_1,y_1,\ldots)+P(x_2,y_2,\ldots)$ and thus for $\lambda=-r_0=\bar{P}(0,0,\ldots)$,
$$\bar{P}(x,y,\ldots)= \lambda + \bar{P}(x_1,y_1,\ldots)+\bar{P}(x_2,y_2,\ldots).$$

%%since  $\bar{P}(x,y,\ldots)= \lambda + \bar{P}(x_1,y_1,\ldots)+\bar{P}(x_2,y_2,\ldots)$ \\
%$m_k(x_i,y_i,\ldots; a_i,b_i,\ldots)=m_k(x_i,y_i,\ldots; a,b,\ldots)$.
%As for polynomials in a commutative ring, non commutative
%polynomials are defined as linear combinations of monomials but the ordering of factors in the monomial
%does matter. Monomials $x.y$ and $y. x$ have to be considered as different and generally, $r.x \neq x.r$ for
%a constant $r\in R$. A polynomial with coefficients in $R$ may be written as
%$P(x,y,z\ldots)=\sum_j a_j m(x,y,z,\ldots)b_j$
%where $m$ is a monomial on the non commuting variables $\{x,y,z\ldots\}$ and $\{a_j,b_j\}$ is a set of
%constants chosen in $R$ taken. It defines a function $P:(x,y,z\ldots) \to P(x,y,z\ldots)$ that maps n-uples of
% $R^n$ to elements of $R$. The problem of such definition is that the product of two polynomial function
% is not necessarily a polynomial function.  Indeed
%Therefore we are leaded to enlarge the definition of polynomial functions to product of polynomials.
%This can be see as ... We This can be generalized
%
%$m(x_1,\ldots,x_k)$ is a finite product of elements taken in the set $\{x_1,\ldots,x_k\}$ of non commuting
%variables.
%different monomials and
% non commutative polynomial functions $R[x_1,\ldots,x_k]=\{\sum_j a_j m(x_1,\ldots,x_k)b_j \}$ where
%$m(x_1,\ldots,x_k)$ is a finite product of elements taken in the set $\{x_1,\ldots,x_k\}$ of non commuting
%variables.

\emph{The Sylow decomposition of a finite ring.}
Let $R$ be a finite ring and let $|R|=\prod_i p_{i}^{e_i}$ be the factorization of its order into powers of
distinct prime numbers. The additive group $R,+$  is a finite commutative group. Hence, it is equal to the
direct product $\oplus_i Syl(p_i)$ of its Sylow subgroups and  thus every element $r$ of $R,+$ can
be written in a unique way as $r=\sum r_i$ where $r_i\in S_{p_i}:=Syl(p_i)$.
 We call the element $r_i$,
\emph{the $p_i$-component of $r$} and it is the unique element contained in the intersection
$S_{p_i}\cap \{r+(\oplus_{j\neq i} S_{p_j})\}$ ( see \cite{Hall}, chapter 3).
These subgroups may be defined as
$S_{p_i}:=\{x : p_{i}^{e_i} x=0\}$ where $p_{i}^{e_i} x$ is the repeated sum of $p_{i}^{e_i}$ terms $x$.
The subgroups $S_{p_i}$ are two-sided ideals of the ring $R$, i.e. $r.S_{p_i}=S_{p_i}.r=S_{p_i}$ for
every $r\in R$. %belong to for $x\in S_{p_i}$  for every $r\in R$.
This is due to the right and left distributive property of a ring since
$p_{i}^{e_i}(r.x)=\underbrace{r.x+\ldots +r.x}_{p_{i}^{e_i} terms}=
r(\underbrace{x+\ldots +x}_{p_{i}^{e_i} terms})= r.(p_{i}^{e_i}x)=r.0=0$ if $x\in S_{p_i}$ so that $r.x\in S_{p_i}$
and similarly $x.r\in S_{p_i}$. Hence every finite ring is the direct sum of its Sylow ideals and a finite
ring that is not decomposable as a non trivial direct sum, must be of prime power order.
Moreover, if $|R|=d_1.d_2$ is the product of two coprime numbers ($\geq 2$) then $R=R_1\oplus R_2$ where
$R_i:=\oplus_{p|d_i}Syl(p).$\\

\emph{Ring with unity.}
From now on, we mainly consider rings $R$ containing a multiplicative unity $1$ such that $x.1=x=1.x$ for
every $x\in R$. If $R=\oplus_{i\in I}R_i$ has a unity $1$ and $R_i\neq \{0\}$ then $1_i$ is the unity of $R_i$.
An element $x$ has a left inverse $x_L$ (resp. right inverse $x_R$) if $x_L .x=1$ (resp $x.x_R =1$). 
An element that has both a left and a right inverse is called a \emph{unit}. If $x$ is a unit, then the inverse 
$x^{-1}$ is unique  since $x_L =x_L.(x.x_R)=(x_L.x).x_R=x_R$. The set $U(R)$ of all units of $R$ is a 
multiplicative group and  by the componentwise multiplication $U(R_1\oplus R_2)=U(R_1)\oplus U(R_2)$.
A \emph{field} is a ring where every non zero element is a unit.
%For finite commutative rings with unity, the Sylow decomposition into rings of prime power order may be
%refined into a sum of \emph{local rings}. A commutative local ring is a ring that has a unique maximal ideal
%$M$. A local ring cannot be decomposed into a non trivial direct sum and hence it is of prime power order.
%  Every finite commutative ring with unity is a direct sum of local rings (see \cite{MacDo}, Proposition %8.7).\\

%%=(x+\ldots +x)=r.(p_{i}^{e_i}x)=r.0=0$ if $x\in S_{p_i}$ so that $r.x\in S_{p_i}$
%%%$p_{i}^{e_i}(r.x)=r.x+\ldots +r.x=r(x+\ldots +x)=r.(p_{i}^{e_i}x)=r.0=0$ if $x\in S_{p_i}$ so that $r.x\in S_{p_i}$
%%$$E(\xi,\varphi_1)\stackrel{(\omega_1,\pi_1)}{\longrightarrow}  E(\xi,\varphi_2)
%%%\stackrel{(\omega_2,\pi_2)}{\longrightarrow} E(\xi,\varphi_3)$$
%%%%$\stackrel {\underbrace{x+\ldots +x}}{\rm {p_{i}^{e_i}}}$
%%%$\underbrace{x+\ldots +x}_{p_{i}^{e_i} terms}$
%\begin{itemize}
%\item unique decomposition : every element can be written in a unique way as $a=\sum a_i$.
%\item every linear map preserves the Sylow ideals.
%\end{itemize}
%\begin{proposition} A finite ring is the direct sum of its Sylow ideals and $(x.y)_i=x_i .y_i$.
%\end{proposition}

\emph{Nilpotency.}
An element $n$ of a ring $R$ is called nilpotent if $n^t=0$ for some positive integer $t$. The set $Nil(R)$
of all nilpotent element of $R$ is called the \emph{Nilpotent radical} of $R$. Once again, by the 
componentwise multiplication $Nil(R_1\oplus R_2)=Nil(R_1)\oplus Nil(R_2)$. We say that an ideal 
$N$ is nilpotent provided that every $n\in N$ is nilpotent.  For every $r\in R$, if we have $(r.n)^t=0$ then
 $(n.r)^{t+1}=n.(r.n)^t .r=0$ and thus every nilpotent ideal is two-sided.
In every ring with unity, a nilpotent element
cannot be a unit but if $n$ is nilpotent ($n^t=0$) then $1 + n$ is a unit. To show this, consider
$u_t (n)=1+n+\ldots + n^{t-1}$;  then since $1=1-n^t=(1 -n)u_t(n)=u_t(n)(1-n)$, % for every $x\in R$,
the element $u_t(-n)$ is the inverse of $1+n=1-(-n)$. \\
%% since $1-x^t=(1 -x)u_t(x)=u_t(x)(1-x)$ for every $x\in R$.\\
%%ùA nilpotent ideal is always two sided. No false
Let us show that in a commutative ring, $Nil(R)$ is an ideal. If $x^n=0$ then $(r.x)^n=r^n.x^n=0$ and if
moreover $y^m=0$ then $(x+y)^{n+m}=0$ since $(x+y)^{n+m}$ is a sum of terms $x^{n+m-k}.y^k$ which are zero for  $k\leq m$ and for $k\geq m$. In a non commutative ring $R$, $Nil(R)$ is not necessarily an ideal. For instance, a sum of nilpotent matrices may be invertible (and thus non nilpotent) as shown by
$${\footnotesize \left( \begin{array}{cc}
0 & 1 \\
1 & 0
\end{array}\right)
=\left( \begin{array}{cc}
0 & 0 \\
1 & 0
\end{array}\right)
+
\left( \begin{array}{cc}
0 & 1 \\
0 & 0
\end{array}\right).}
$$
Thus for the ring $M_2(R)$ of $2\times 2$ matrices over a ring $R$ with unity, the nilpotent radical is not an ideal. However the subring of upper triangular matrices is also non commutative but it contains the nilpotent ideal $\{\scriptsize {\left( \begin{array}{cc}
0 & r \\
0 & 0
\end{array}\right)} :r\in R\}.$

%A local ring cannot be decomposed into a non trivial direct sum and hence it is of prime power order.

\subsection{Generalizing $Teichm \ddot{u} ller$ sets}\label{teichmull}\label{Teich}
For every ring $R$ we would like to define a subset $S_R\subset R$ for the indexes $k,l$ of vectors and
components, in such a way that for a Galois ring $R=GR(4,n)$, the set $S_R$ is the $Teichm \ddot{u} ller$
set $T_n$ as in formula (\ref{galois}), while for a finite field $R=\mathbb{F}_{p^n}$ we have $S_R=R$ as in
formula (\ref{pnodd}).
%% If $N$ is an ideal of a ring $R$ Then
% As explained previously, in a commutative ring $R$, the set $Nil(R)$ of all nilpotent elements is an ideal.

The set $T_n \subset GR(4,n)$ has remarkable properties that are used over
and over to compute  easily in $GR(4,n)$ (see \cite{MacDo2}).
\begin{enumerate}
\item The ideal $N:=\{2t_1:\,\,t_1\in T_n\}$ is the nilpotent radical of $GR(4,n)$~($N^2=~0$).
\item Every $r\in GR(4,n)$ can be written in a unique way as $r=t_o +2t_1$ for some $t_0,t_1$ in $T_n$.
Thus $T_n$ contains exactly one representative of each coset $\{r+N\}_{r\in R}$ of $N$ in $R$ ; it is a transversal to the ideal
$N$.
\item $T_n$ is closed under multiplication. Therefore a product of elements written as
$t +n$ for $t\in T_n$ and $n\in N$ is still written in this way since
$(t_1 +n_1)(t_2 +n_2)=\underbrace{t_1t_2}_{\in T_n} +(\underbrace{t_1n_2 +n_1t_2 +n_1n_2}_{\in N}).$
\end{enumerate}
This may be generalized to every ring $R$ as follows. We require that the set of indexes $S_R$ is
closed under multiplication and that it is a transversal to a nilpotent ideal $N$.
Trivially, if $R$ is a field (as in formula (\ref{pnodd})) or even a division ring, then $N=\{0\}$ is the only nilpotent ideal, $S_R=R$ is the only transversal to $N$ and it is closed under multiplication.\\

A commutative local ring is a ring that has a unique maximal ideal $M$ and the Galois ring $R=GR(4,n)$ is
local. In a finite commutative local ring $R$, the unique maximal ideal is $Nil(R)$ and the units of $R$ are exactly the non nilpotent elements (see \cite{MacDo}). Hence in a local ring every ideal ($\neq R$) is nilpotent. Every finite commutative ring with unity is a direct sum of local rings (see \cite{MacDo}, Proposition 8.7).\\
%First show that this may only be defined on a local ring since in general there is no multiplicative %subgroup that is  a transversal to the nilpotent.
%
%Since in the non commutative case $Nil(R)$ is not
%necessarily an ideal, we only require that $S$ is a transversal to a nilpotent ideal $N$ ( may be smaller
%than $Nil(R)$).
%
%uniqueness of writting is due do the fact that $M$ is an additive subgroup and the multiplication of cosets
%is due to the ideal property.  Closed under mult gives the correct writting for  a product of element.
%A good reference is \cite{MacDo2}. DEFINE A \emph{transversal}.

%Proof.  Every finite commutative ring $R$ is a sum $R=\oplus_i L_i$ of finite commutative local ring $L_i$.
%Since each $L_i$ is local, $Nil(L_i)$ is maximal whence $R/L_i$ is a finite field and thus the multiplicative
%group of $R/L_i$ is cyclic. to show that $R$ has an element of multiplicative order $R/Nil -1$ i need
%schur-zassenhaus.
%$S_R$ is closed under multiplication as a product of sets closed under multiplication.
%It is a transversal to $\oplus_i Nil(L_i)$ which is equal to $Nil(R)$.

\subsection{Functions preserving a direct sum decomposition}\label{fpres}
Let $R=R_1\oplus R_2$ be a direct sum decomposition of a ring $R$ and let $r=r_1 +r_2$
be the corresponding decomposition for $r\in R$.
Let $G,_{\star}$  be a commutative group with an operation $\star$ ( either $"+"$ or $"\cdot"$ in this paper).
For a finite set of variables $\{x,y,\ldots\}$ belonging to $R$, we say that a function
$f:(x,y,\ldots)\to f(x,y,\ldots)\in G,_{\star}$ \emph{preserves the decomposition $R_1\oplus R_2$} if for
a constant $\lambda\in R$
% $$f(x,y,z,\ldots)=f(x_1,y_1,z_1,\ldots)\star f(x_2,y_2,z_2,\ldots)\quad \textrm{for every}\quad  x,y,z,\ldots\in R$$.
\begin{equation}\label{fpres}
 f(x,y,\ldots)=\lambda\star f(x_1,y_1,\ldots)\star f(x_2,y_2,\ldots)\quad
 \textrm{for every}\quad  x,y,\ldots\in R.
 \end{equation}
Observe that $\lambda =(f(0,0,\ldots))^{-1}$ because $(x_1)_2=(y_1)_2=\ldots=0$ implies
$f(x_1,y_1,\ldots)=f(x_1+0,y_1+0,\ldots)=\lambda\star f(x_1,y_1,\ldots)\star f(0,0,\ldots).$
If $f_i$ is the restriction of $f$ to $R_i$ then
$f(x,y,\ldots)=\lambda\star f_1(x_1,y_1,\ldots)\star f_2(x_2,y_2,\ldots)$. Conversely,
 for arbitrary functions $f_i$ from $R_i$ into $G$ and $\lambda\in G$, this last equation defines a function
that preserves $R_1\oplus R_2$. It may happen that $f$ preserves $R_1\oplus R_2$ but does not preserve another decomposition of $R$.\\

We have seen in subsection \ref{finite rings} that polynomial functions $P$ on a ring $R$ preserves every direct sum
decomposition of $R$ ( and in this case $\lambda=-P(0,0,\ldots)$).
Thus, since group homomorphisms preserve sums, if $T$ is a group homomorphism from
$R_{+}$ into a commutative group $G_{\star}$ and if $P(x,y,\ldots)$ is a polynomial function on $R$, then
$(x,y,\ldots)\to T(P(x,y,\ldots))$ preserves every direct sum decomposition of $R$.
Hence, these functions generalize formula (\ref{pnodd}) and (\ref{galois}) for mutually unbiased bases
since those rely on expressions of type
$\frac{1}{\sqrt{N}} T (P (k,l))$ for $k,l\,\in R$ where $G_{\star}$ is the multiplicative group of unitary
complex number. This is also true for the other formulas proposed in \cite{RotKlap}.\\

%Let $G,\cdot$ be a group and let $F:(x_1,\ldots,x_k)\to F(x_1,\ldots,x_k)\in G$ be a
%k-variables function into $G$ with variables $x_1,\ldots,x_k$ in $R$. We say that the function $F$
%\emph{preserves the Sylow decomposition of $R,+$} if
%$$F(x_1,\ldots,x_k)=\prod_{p_i} F((x_1)_i,\ldots,(x_k)_i)$$
%where products in the second member are performed into $G,\cdot$.\\
%%%%%%%%%%%%%%%%%%%%%%%%%%%%%%%%%%%%%%%%%%%%%%%%%%%

%\subsubsection{Functions preserving the Sylow decomposition}
%Let $R$ be a finite ring, let $\{p_i\}$ be the distinct prime factors of $|R|$ and for $r\in r$, let
%$r_i$ be the $p_i$-component of $r$. Then $R$ is the ring direct sum $\oplus_i R_i$.

%Let $G,\cdot$ be a group and let $F:(x_1,\ldots,x_k)\to F(x_1,\ldots,x_k)\in G$ be a
%k-variables function into $G$ with variables $x_1,\ldots,x_k$ in $R$. We say that the function $F$
%\emph{preserves the Sylow decomposition of $R,+$} if
%$$F(x_1,\ldots,x_k)=\prod_{p_i} F((x_1)_i,\ldots,(x_k)_i)$$
%where products in the second member are performed into $G,\cdot$.\\

More sophisticated such functions may be constructed by products. If $f\cdot g$ is a product of functions into a commutative group $G_{\cdot}$ that both preserve a direct sum decomposition $R=R_1\oplus R_2$ then  it is easy to show that $f\cdot g$ also preserves $R_1\oplus R_2$. We have
$(f\cdot g) (x,y,\ldots)=\lambda_f\cdot\prod_{i=1,2}   f(x_i,y_i,\ldots)\cdot
\lambda_g\prod_{i=1,2}  g(x_i,y_i,\ldots)$ and since the elements of $G$ commute,
we may rearrange the factors as $(f\cdot g) (x,y,\ldots)=\lambda_f\cdot \lambda_g \prod_{i=1,2}
(f\cdot g)(x_i,y_i,\ldots).$

\section{Such sets of MUB cannot be complete}
%%\section{There are at least $1+min(p_1^{e_1},\,i=1\ldots r)$ bases}\label{lowerbound}
In this section we prove that even with all these generalizations, it is not possible to construct complete
sets of $N+1$ mutually unbiased bases for $N\neq p^n.$

\subsection{Preliminary results}
%%%%%%%%%%%%%%%%%%%%%%%%%%%%%%%%%%%ù
\begin{proposition} \label{TeichSum}
In a ring $R$ with 1,
let $S\subset R$ be a set closed under multiplication that is a transversal to a nilpotent ideal $N$ of $R$.
\begin{enumerate}
\item If $R=R_1\oplus R_2$ is a sum of rings with $1$ then
$$ S=(S\cap R_1)\oplus  (S\cap R_2)$$ and each $S\cap R_i$ contains at least two elements ($i=1,2$).
 Moreover in each $R_i$, $S\cap R_i$ is closed under multiplication and is a transversal
 to $N\cap R_i$.
\item If $|S|$ is a product $d_1.d_2$ of two coprime numbers $\geq 2$ and if $R$ is finite, then $R$ is a sum $R_1\oplus R_2$ of rings with
 $1$ such that $|S\cap R_1|=d_1$ and $|S\cap R_2|=d_2.$ %% there is a decomposition $R=R_1\oplus R_2$
\end{enumerate}
%Let $R=R_1\oplus R_2$ be a sum of rings with $1$.
%Let $S\subset R$ be a set closed under multiplication that is a transversal to a nilpotent ideal $N$ of $R$.
% Then
% $$ S=(S\cap R_1)\oplus  (S\cap R_2)$$ and each $S\cap R_i$ contains at least two elements ($i=1,2$).
\end{proposition}
\emph{Proof.}
1) $R=R_1\oplus R_2$ has unity $(1_1,1_2)$. First, we show that every ideal $I$ of $R$ is equal to $I_1\oplus I_2$
 where $I_j$ is an ideal of $R_j$. Since $I$ is an ideal of $R$, the sets
 $I_1:=I.(1_1,0)$ and $I_2:=I.(0,1_2)$ belong to $I$, are ideals of $R$ and thus in particular $I_j$ is an ideal
 of  $R_j$. But since every $i=(i_1,i_2)\in I$ is equal to $i.(1_1,0)+i.(0,1_2)$, we have $I=I_1\oplus I_2$.
 By componentwise multiplication, an element $(n_1,n_2)$ is nilpotent if and only if each $n_j$ is nilpotent
 in $R_j$ and the ideal $N$ is the sum $N_1\oplus N_2$ of two nilpotent ideals. Since $S$ is a transversal
 to $N$ in $R$, it contains a unique element of each coset of $N$.  Let $x$ be the unique element
 $S\cap \{(1_1,0)+N\}$ then $x=(1_1+n_1,n_2)$ for some $(n_1,n_2)\in N_1\oplus N_2$.
 Since $N$ is a nilpotent ideal,it is two-sided  and  we may consider the quotient ring $R/N$ where multiplication of cosets is defined
 as $(x+N)(y+N)=x.y +N$. In $R/N$, $((1_1,0)+N)^2=(1_1,0)+N$ so that $x^2 \in (1_1,0)+N$ and also
 $x^2 \in S$  because $S$  is closed under multiplication.
 Therefore we have $x^2=S\cap \{(1_1,0)+N\}=x$ and we have

$$ (1_1 + n_1)^2=(1_1 + n_1) \quad (1),\quad n_2^2 =n_2\quad (2). $$
By nilpotency $n_2^t=0$ for some positive integer $t$ and by $(2)$, $n_2^t=n_2$ whence $n_2=0$.
By nilpotency of $n_1$, $(1+n_1)$ has an inverse $(1_1+n_1)^{-1}$ in $R_1$. Multiplying both sides of $(1)$
 by $(1_1+n_1)^{-1}$ gives $(1_1+n_1)=1$ whence $n_1 =0$. Finally $x=(1_1,0)\in S$ and the symmetric
  argument for $S\cap \{(0,1_2)+N\}$ shows that $(0,1_2)$ also belongs to $S$. Thus
  $(0,0)=(1_1,0).(0,1_2)\in S$,  and  $S\cap R_1$ (resp. $S\cap R_2$) contains at least the two elements
  $(0,0)$ and $(1_1,0)$ (resp.  $(0,0)$ and $(0,1_2)$).

As $\{(1_1,0),(0,1_2)\}\in S$, for every $(s_1,s_2)\in S$, $(s_1,0)=s.(1_1,0)\in S\cap R_1$ and
$(0,s_2)=s.(0,1_2)\in S\cap R_2$. Conversely, it remains to show that for every  $(s_1,0)\in S\cap R_1$ and
 $(0,s_2)\in S\cap R_2$ we also have $(s_1,s_2)\in S$. Since $S$ is a transversal to $N$, it contains a unique
 element $y=(s_1+n_1,s_2+n_2)\in S\cap \{(s_1,s_2)+N\}$ and $y.(1_1,0)=(s_1+n_1,0)$ in $S$.
As $(s_1+n_1,0)$ and $(s_1,0)$ are in $S$ and  belong the same coset of $N$, these must be equal and
$n_1=0$. Similarly $n_2=0$ so that $y=(s_1,s_2)\in S$ whence $ S=(S\cap R_1)\oplus  (S\cap R_2)$.
Finally, let us show that $(S\cap R_i)$ is a transversal to $N_i=N\cap R_i$ in $R_i$.
 Every coset $(r_1,0)+N_1$ is embedded in $(r_1,0)+N$ which contains a unique element
 $s=(r_1+n_1,n_2)$ of $S$. Then $s.(1_1,0)=(r_1+n_1,0)$ is in $S\cap R_1$ and in $(r_1,0)+N_1$ and so,
 $S\cap R_1$ contains at least one representative of each coset.
If two elements of $S\cap R_1$ belong to the same coset of $N_1$ then these belong to the same coset of 
$N\supset N_1$ and thus are equal. The proof is similar for $(S\cap R_2)$.\\

2) Let $\pi(d)$ denote the set of prime divisors of a positive integer $d$. For a finite ring $R$ and a
divisor $d$ of $|R|$, let us define $Syl(\pi(d)):=\oplus_{p\in\pi(d)}Syl(p)$. Since $N$ is a transversal to
the ideal $N$ in $R$ then $d_1.d_2=|S|=\frac{|R|}{|N|}$ is a divisor of $|R|$. If $\pi_1=\pi(d_1)$ and
$\pi_2=\pi(|R|)\setminus\pi(d_1)$, we know from subsection \ref{finite rings} that
$R=Syl(\pi_1)\oplus Syl(\pi_2).$ For $i=1,2$, since $d_i\geq 2$, the subsets $\pi_(d_i)$ are non empty
so that $\pi_1=\pi(d_1)$ and $\pi_2\supset \pi(d_2)$ are non empty ; whence the rings $Syl(\pi_i)$ are not
zero rings and since $R$ has unity $1$, $Syl(\pi_i)$ has unity $1_i$. Thus part (1) of the present
 proposition applies, $S=(S\cap Syl(\pi_1))\oplus (S\cap Syl(\pi_2))$ and
\begin{equation}\label{s_syl}
 d_1.d_2=|S|= |S\cap Syl(\pi_1)|.|S\cap Syl(\pi_2)|.
 \end{equation}
As $S\cap Syl(\pi_i)$ is  a transversal to the ideal $N\cap Syl(\pi_i)$ then $|S\cap Syl(\pi_i)|$ divides
$Syl(\pi_i)$ and so is coprime to $d_j$ ($j\neq i$). Thus, by equality (\ref{s_syl}), $|S\cap Syl(\pi_i)|$
must divides $d_i$ and symmetrically $d_i$ divides $|S\cap Syl(\pi_i)|$ so that $d_i=|S\cap Syl(\pi_i)|.$

%To conclude, it is sufficient to notice that if $d$ divides a product $C_1.C_2$ of coprime
%numbers but $d$ is coprime with $C_1$, then $d$ must divide $C_2$. Thus as $N_1$ divides
%$|S\cap R_1|$ and $|S\cap R_1|$ divides $N_1$, we have $N_1=|S\cap R_1|$ and
%similarily $N_2=|S\cap R_2|.$
 $\quad\Box$\\

%If $R$ is a finite commutative ring with $1$, then there exists a transversal to $Nil(R)$ that is closed
% under multiplication.
%%%%%%%%%%%%%%%%%%%%%%%%%%%%%%%%%%%%%%%%
In what follows,  $\langle  | \rangle $ denotes the classical hermitian product $\langle a,b \rangle =\sum_{i}(a_i)^{*}b_i$.
%so that $\langle \lambda a|b \rangle =\lambda^{*}\langle a|b \rangle $ for a complex number $\lambda$ and  $a$, $b$ complex vectors. %(convention pg 373 Wooters and Fields)
The tensor product $v\otimes w$ is defined by $(v\otimes w)_{(i,j)}=a_i.b_j$ and thus
 $\langle v_1\otimes w_1 | v_2\otimes w_2 \rangle =\langle v_1|v_2 \rangle .\langle w_1|w_2 \rangle $.

% An irreducible summand is a  pregeometry that is not a direct sum
%and we call such a $\Gamma$ an \mbox{\boldmath$irreducible\ pregeometry$}.\\

\begin{proposition}\label{splitmub}
Let $N=N_1.N_2$ be a product of positive integers and
let $\{v_{i,j}^{(1)}\}\ldots \{v_{i,j}^{(r)}\}$ be $r$ mutually unbiased bases of
$\mathbb{C}^N \cong \mathbb{C}^{N_1}\otimes \mathbb{C}^{N_2}$ (for $(i,j)\in \bar{N}:=\{1\ldots N_1\}\times \{1\ldots N_2\})$. Assume that for each $1\leq t\leq r$ there are
$N_1$ vectors $\{a_i^{(t)}\}$ and  $N_2$ vectors $\{b_j^{(t)}\}$ such that $$v_{i,j}^{(t)}=a_i^{(t)}\otimes  b_j^{(t)}$$ % \qquad \textrm{for every } (i,j)\in \bar{N}. $$
for every $(i,j)\in \bar{N}$, then
%%\begin{enumerate}
%%\item
for $1\leq t\leq r$, $\{\frac{a_i^{(t)}}{||a_i^{(t)} ||}\}$ and $\{\frac{b_j^{(t)}}{||b_j^{(t)}||}\}$ are r mutually unbiased bases respectively in $\mathbb{C}^{N_1}$ and $\mathbb{C}^{N_2}$.
%%\item If moreover $\{v_{i,j}^{(t)}\}$ is unbiased with the standard basis  of $\mathbb{C}^{N}$ then
%$\{\frac{a_i^{(t)}}{||a_i^{(t)} ||}\}$ (resp. $\{\frac{b_j^{(t)}}{||b_j^{(t)}||}\}$) is unbiased with the %%standard basis $\mathbb{C}^{N_1}$ (resp.$\mathbb{C}^{N_2}$ ).
%%\end{enumerate}
\end{proposition}
\emph{Proof.}
First we show that $\{a_i^{(t)}\}$ and  $\{b_j^{(t)}\}$ are orthogonal bases. Since othonormality of 
$\{v_{i,j}^{(t)}\}$ implies
$\langle  v_{i,j}^{(t)}|v_{k,l}^{(t)} \rangle =\delta _{\{(i,j),(k,l)\}}=\delta _{ik}\delta _{jl}$, we have $\delta _{ik}\delta _{jl}=\langle a_i^{(t)}\otimes b_j^{(t)}| a_k^{(t)}\otimes b_l^{(t)} \rangle =\langle a_i^{(t)}|a_k^{(t)} \rangle \langle b_j^{(t)}|b_l^{(t)} \rangle $. For every $j$, $b_j^{(t)}\neq \bar{0}$  %\mathbb{o}$
 (otherwise $v_{i,j}^{(t)}=\bar{0}$) %%%\mathbb{0}$)
 so that $\langle a_i^{(t)}|a_k^{(t)} \rangle \langle b_j^{(t)}|b_j^{(t)} \rangle =\delta_{ik}\delta_{jj}=\delta_{ik}$ implies that $\langle a_i^{(t)}|a_k^{(t)} \rangle =0$ for $i\neq k$. Hence $\{a_i^{(t)}\}$
is a set of $N_1$ mutually orthogonal vectors, thus an orthogonal basis ( not necessarily orthonormal ) of
$\mathbb{C}^{N_1}$.
Permuting the role of $a$ and $b$ gives the same result for $\{b_j^{(t)}\}$ in $\mathbb{C}^{N_2}$.\\
Furthermore if we fix $1\leq j\leq N_2$, the equalities
$1=\langle  v_{i,j}^{(t)}|v_{i,j}^{(t)} \rangle =\langle a_i^{(t)}|a_i^{(t)} \rangle \langle b_j^{(t)}|b_j^{(t)} \rangle $ may be divided by the constant
$\langle b_j^{(t)}|b_j^{(t)} \rangle $ so that for every $1\leq i\leq N_1$, $L_{a^{(t)}}:=\langle a_i^{(t)}|a_i^{(t)} \rangle $ is constant
with respect to $i$
and equal to $1/\langle b_j^{(t)}|b_j^{(t)} \rangle $. Symmetrically, $L_{b^{(t)}}:=\langle b_j^{(t)}|b_j^{(t)} \rangle $ is also constant for every
$1\leq j\leq N_2$ and
%for every $1\leq j\leq N_2$
\begin{equation}\label{LaLb}
L_{a^{(t)}}:=\langle a_i^{(t)}|a_i^{(t)} \rangle =\frac{1}{\langle b_j^{(t)}|b_j^{(t)} \rangle }=\frac{1}{L_{b^{(t)}}}\quad \textrm{for every}\quad (i,j)\in\bar{N}.
\end{equation}

Now, it is sufficient to prove the MUB property for each couple of bases among the r, in $\mathbb{C}^{N_1}$
 and $\mathbb{C}^{N_1}$. For instance let us
consider $\{v_{i,j}^{(1)}\}$ and $\{v_{k,l}^{(2)}\}$. We define $A_{i,k}:=|\langle a_i^{(1)}|a_k^{(2)} \rangle |$ and
$B_{j,l}:=|\langle b_j^{(1)}|b_l^{(2)} \rangle |$. The equality
\begin{equation}\label{mubAB}
1/\sqrt{N}=|\langle \underbrace{v_{i,j}^{(1)}}_{a_i^{(1)}\otimes b_j^{(1)}}|\underbrace{v_{k,l}^{(2)}}_{a_k^{(2)}\otimes b_l^{(2)}} \rangle |=A_{i,k}B_{j,l}\quad\textrm{for every } (i,j),(k,l)\in \bar{N}
\end{equation}
implies that $A_{i,k}$ and $B_{j,l}$ are non zero. Therefore if $(i,k)$ is fixed and $(j,l)$ varies,
$A_{i,k}$ can be simplified and all the $B_{j,l}$ are equal to a common value $K_B$. Symmetrically, the 
$A_{i,k}$ are equal to a common value $K_A$.\\
% and by equality (\ref{mubAB}) we have $K_AK_B=1/\sqrt{N}$.

In basis $\{a_i^{(1)}\}$ we have $a_k^{(2)}=\sum_i\lambda_i a_i^{(1)}$ for
$\lambda_i=\frac{\langle a_i^{(1)} | a_k^{(2)} \rangle }{\langle a_i^{(1)}|a_i^{(1)} \rangle }$. % \\
Now, equality (\ref{LaLb}) and (\ref{mubAB}) prove that $|\lambda_i|=\frac{K_A}{L_{a^{(1)}}}$
 whence it is constant for every $(i,k)$. Therefore $$L_{a^{(2)}}=|\langle a_k^{(2)}|a_k^{(2)} \rangle |=|\langle \sum_i\lambda_i a_i^{(1)}|\sum_{i'}\lambda_{i'} a_{i'}^{(1)} \rangle |$$
 $=\sum_i ^{N_1} |\lambda_i|^2 |\langle a_i^{(1)}|a_i^{(1)} \rangle |=N_1|\lambda_i|^2 L_{a^{(1)}}=N_1(\frac{K_A}{L_{a^{(1)}}})^2 L_{a^{(1)}}=\frac{N_1(K_A)^2}{L_{a^{(1)}}}.$

\begin{equation}\label{linksAK}
\textrm{that is}\quad L_{a^{(1)}}L_{a^{(2)}}=N_1(K_A)^2
%%\quad \textrm{and}\quad L_{b^{(1)}}L_{b^{(2)}}=N_2(K_B)^2
\end{equation}

 Finally, we show that $\{\frac{a_i^{(1)}}{||a_i^{(1)} ||}\}$ and $\{\frac{a_k^{(2)}}{||a_k^{(2)}||}\}$ are mutually unbiased bases  in $\mathbb{C}^{N_1}$. Indeed
 $|\langle \frac{a_i^{(1)}}{||a_i^{(1)}||}|\frac{a_k^{(2)}}{||a_k^{(2)}||} \rangle |^2=
\frac{|\langle a_i^{(1)}|a_k^{(2)} \rangle |^2}{||a_i^{(1)}||^2||a_k^{(2)}||^2}
 =\frac{(K_A)^2}{L_{a^{(1)}}L_{a^{(2)}}}=\frac{1}{N_1}$ (by (\ref{linksAK})).

 The result for $\mathbb{C}^{N_2}$ is obtained in the same way, using $b_l^{(2)}=\sum_j\mu_j b_j^{(1)}$ for
  $\mu_j=\frac{\langle b_j^{(1)} | b_l^{(2)} \rangle }{\langle b_j^{(1)}|b_j^{(1)} \rangle }$, to give
$L_{b^{(1)}}L_{b^{(2)}}=N_2(K_B)^2$. $\quad\Box$\\

%% is constant for every $(i,k)$ and similarily %%$|\langle \frac{b_j^{(1)}}{||b_j^{(1)}||}|\frac{b_l^{(2)}}{||b_l^{(2)}||} \rangle |=\f%%rac{K_B}{\sqrt{L_{b^{(1)}}}\sqrt{L_{b^{(2)}}}}
%%$ is constant for every $(j,l)$.
%%$\stackrel{\beta}{\underbrace{abc}_{rtg}}$

This proposition can immediately be extended as follows to
$\mathbb{C}^N \cong \mathbb{C}^{N_1}\otimes\ldots \otimes \mathbb{C}^{N_s}$ for dimension
$N=N_1\ldots N_s$. Under assumption that each of the k bases is a tensor product, we may use induction
to conclude to the existence of $k$ mutually unbiased bases in each $\mathbb{C}^{N_i}$.

%\begin{theorem}There exists a complete set of $d+1$ mutually unbiased bases described by a
%Wooters \& Fields formula if and only if $d$ is a prime power.
%\end{theorem}
%\begin{corollary}\label{atmost} A Wooters \& Fields formula on a ring of $d$ elements provides at most
%$1+min_{d}(p_1^{e_1},\,i=1\ldots r)$ mutually unbiased bases.
%\end{corollary}
%%%%%%%%%%%%%%%%%%%%%%%%%%%%%%%%%%%%%%%%%%
\subsection{Main results}
\begin{theorem}\label{notN+1}
Let $R=R_1\oplus R_2$ be a decomposition of a ring $R$. For $i=1,2$ let $S_i$ be a non empty subset
of $R_i$ and let $N=|S_1||S_2|$. For each $1\leq c\leq m$, let $f_c:R_{+}\to \mathbb{C}^{*},\cdot$
be a two variables function that preserves the decomposition $R_1\oplus R_2$ and let us define $N$
vectors $\{v^{(c)}_k\}$ of $\mathbb{C}^{N}$ as
$$(v^{(c)}_k)_l = f_c (k,l) \qquad k,l\,\in S_1\oplus S_2.$$ %%% \frac{1}{\sqrt{N}}

Assume that, together with the standard basis, the sets of vectors $\{v^{(c)}_k\}_{1\leq c\leq m}$
form a set $X$ of $m+1$ mutually unbiased bases. If $|S_i|_{i=1,2} \neq 1$, then
\begin{center}
$m\leq \underset{i}{min} |S_i|<N$ and the set $X$ is not complete.
\end{center}
%%\item If $m=N$ then
%\end{center}
\end{theorem}
\emph{Proof.}
The function $f_c$ preserves the decomposition $R_1\oplus R_2$ so
there is a constant $\lambda_c$ such that $f_c(k,l)=\lambda_c f_c(k_1,l_1)f_c(k_2,l_2)$.
For each $c$,  let us define $|S_1|$ vectors $\{a^{(c)}_{k_1}\}_{k_1\in S_1}$ of
$\mathbb{C}^{|S_1|}$ and $|S_2|$ vectors $\{b^{(c)}_{k_2}\}_{k_2\in S_2}$ of $\mathbb{C}^{|S_2|}$ as
$$(a^{(c)}_{k_1})_{l_1} =\lambda_c f_c(k_1,l_1) \quad , \quad
(b^{(c)}_{k_2})_{l_2} =f_c(k_2,l_2) \quad for \quad k_i,l_i \in S_i \, (i=1,2).$$
Hence $(v^{(c)}_k)_l=f_c(k,l)=\lambda_c (a^{(c)}_{k_1})_{l_1}(b^{(c)}_{k_2})_{l_2}$ and since $l$ takes all
value in $S_1\oplus S_2$, the vector $v^{(c)}_k=v^{(c)}_{(k_1,k_2)}$ is equal to the tensor product
$a^{(c)}_{k_1}\otimes b^{(c)}_{k_2}\in \mathbb{C}^{|S_1||S_2|}.$
 If we denote by $\{v^{(0)}_k\}_{k\in S_1\oplus S_2}$, $\{a^{(0)}_{k_1}\}_{k_1 \in S_1}$,
  $\{b^{(0)}_{k_2}\}_{k_2 \in S_2}$ the standard bases respectively in
$\mathbb{C}^{N}$, $\mathbb{C}^{|S_1|}$ and $\mathbb{C}^{|S_2|}$ then also
$v^{(0)}_{k}=v^{(0)}_{(k_1,k_2)}=a^{(0)}_{k_1}\otimes b^{(0)}_{k_2}$.

Therefore if the sets $\{v^{(c)}_k\}_{0\leq c\leq m}$ form %%%together with the standard basis,
a set $X$ of $m+1$ mutually unbiased bases in $\mathbb{C}^{N}=\mathbb{C}^{|S_1||S_2|}$, then by Proposition \ref{splitmub},
 there exist $m+1$ mutually unbiased bases in both
$\mathbb{C}^{|S_1|}$ and $\mathbb{C}^{|S_2|}$. By Theorem \ref{WoFi}, if each $|S_i|_{i=1,2}$ is at least
$2$ then $m+1\leq |S_i|+1$ and $m\leq \underset{i}{min} |S_i|<|S_1||S_2|=N$ thus $|X|=m+1<N+1$ and $X$ is not complete.
$\quad\Box$\\
%%%%%%%%%%%%%%%%%%%%%%%%%%%%%%%%%%%%%%%%%%%
%%%%%%%%%%%%%%%%%%%%%%%%%%%%%%%%%%%%%%%%%%

Finally, we obtain our main result : complete sets of MUB described by generalizations of known formulas
only exist for prime power dimensions. Moreover, we provide an upper bound for the number of MUB described by such formulas.

%MMMMMMMMMMMMMM
%\begin{theorem}
%Let $R$ be a finite ring of order $N$ with unity.
%Let $S$ be a subset of $R$ closed under multiplication and transversal to a nilpotent ideal.
%For $1\leq c\leq N$, let $T_c:R_{+}\to \mathbb{C}^{*},\cdot$
%be a group homomorphism and let $P_{c}:R^2 \to R$ be a two variables polynomial function.
%Assume that, together with the standard basis, the sets of vectors $\{v^{(c)}_k\}_{1\leq c\leq N}$ of
% $\mathbb{C}^{N}$ defined by $$(v^{(c)}_k)_l =\frac{1}{\sqrt{N}} T_c (P_c (k,l)) \qquad k,l\,\in S$$
%form a set of $N+1$ mutually unbiased bases. Then $N$ is a prime power.
%\end{theorem}
\begin{theorem}
Let $R$ be a finite ring with unity.
Let $S\subset R$ be a subset of $N$ elements that is closed under multiplication and transversal to a nilpotent ideal.
For $1\leq c\leq N$, let $T_c:R_{+}\to \mathbb{C}^{*},\cdot$
be a group homomorphism and let $P_{c}:R^2 \to R$ be a two variables polynomial function.
Let us define $N$ sets of vectors $\{v^{(c)}_k\}_{1\leq c\leq N}$  in $\mathbb{C}^{N}$ by
$$(v^{(c)}_k)_l =\frac{1}{\sqrt{N}} T_c (P_c (k,l)) \qquad k,l\,\in S,$$
and let the set $X_{\{T_c\}\{P_c\}}$ be the union of the standard basis with $\{v^{(c)}_k\}_{1\leq c\leq N}.$

\begin{enumerate}
\item A set $X_{\{T_c\}\{P_c\}}$ contains at most $1+\underset{i}{min}\{p_i ^{e_i}\}$ mutually unbiased bases where
$N=\prod_i p_{i}^{e_i}$ is the factorization of $N$ into powers of distinct prime numbers.

\item There exists a complete set $X_{\{T_c\}\{P_c\}}$ of $N+1$ mutually unbiased bases if and only if
$N$ is a power of a prime number.

\end{enumerate}
\end{theorem}
\emph{Proof.}
\begin{enumerate}
\item
We prove that the conditions of Theorem \ref{notN+1} applies here.\\
First we show that every vector $v^{(c)}_k$ is unbiased with the standard basis $\{e_k\}$.
% As $T_c:R_{+}\to \mathbb{C}^{*},\cdot$ is a group homomorphism and $R_{+}$ is a finite group,
%its image $T_c(R)$ is a finite subgroup of$\mathbb{C}^{*},\cdot.$
Every $r\in R$ has finite additive order $n_r$ ($n_r.r=0$).
By the homomorphism property $T_c(n.r)=T_c(\underbrace{r+\ldots +r}_{n\,\, terms})=(T_c(r))^{n}$
and as $T_c(0)=1$ we must have $|T_c(r)|^n=1$ whence $|T_c(r)|=1$ for every $r$ in $R$.
Thus for every vector $v^{(c)}_k$ we obtain %the MUB property is satisfied since
 $|\langle e_l|v^{(c)}_k \rangle |=|(v^{(c)}_k )_l|=|T_c (P_c (k,l))|/\sqrt{N}=1/\sqrt{N}$ as announced.
Let $m+1$ be the maximal number of mutually unbiased bases contained in $X_{\{T_c\}\{P_c\}}$ and let
$Y\subset X_{\{T_c\}\{P_c\}}$ be a set of $m+1$ mutually unbiased bases. Since we have showed that
$Y\cup \{\{e_k\}\}$ is also a set of MUB, the standard basis $\{e_k\}$ must be in $Y$.\\

As $|S|=N=\prod_i p_{i}^{e_i}$ we may use Proposition \ref{TeichSum} (2) to show that
there is a ring decomposition
$R=\oplus_i R_i$  such that $S=\oplus_i S\cap R_i$ and $|S\cap R_i|=p_{i}^{e_i}.$
Finally, since the functions $(k,l)\to T_c (P_c (k,l))$ preserve every direct sum decomposition of $R$
(see subsection \ref{fpres}), we may apply Theorem \ref{notN+1} to $Y$ to show that for every $i$ we must have $m\leq p_{i}^{e_i}$. Hence $1+m\leq 1+ \underset{i}{min}\{ p_{i}^{e_i}\}.$

\item If we have $N+1$ such MUB, then by (1), $\prod_i p_{i}^{e_i}=N\leq \underset{i}{min} \{p_{i}^{e_i}\}$,
 which implies that $N=\underset{i}{min} \{p_{i}^{e_i}\}$ and thus $N$ is a prime power.
  Conversely if $N=p_{i}^{e_i}$, we have shown in section \ref{formulas} that the sets of
$N+1$ MUB given by formulas (\ref{pnodd}) and (\ref{galois}) may be described as sets $X_{\{T_c\}\{P_c\}}$

 \end{enumerate}
$\quad\Box$\\
%%%%%%%%%%%%%%%%%%%%%%%%%%%%%%%%%%%%%%%%%%%
%\subsection{consequences : formulas for direct products of finite fields.}
%Ivanovic's formula \ref{ivanovic} makes sense for an arbitrary integer $N$ ; addition and multiplication in
%the exponents are then performed in the ring of integers modulo $N$. The question is to know whether
%these $N$-dimensionnal vectors describe sets of MUB. In order to generalize Wooters \& Fields formula
%\ref{Woot} (for an odd prime power) we need to define a ring $A_N$ with $N$ elements and a Trace
%function that maps elements of $A_N$ into a cyclic subgroup $Z_{\tilde{N}}$. A way to achieve this is the
%following : there is an isomorphism from $Z_{\tilde{N}}$ into the direct product ... described by ...
%A solution to generalize \ref{Woot} is to choose $A_N$ to be the direct product of the finites fields $p^{e}$
%where the factorization of $N$ is ... The trace function on each components induce a trace on the product.
%This Trace function has the same "linearity" properties as the original ones.\\
%
%Let us show that this generalization leads to a set of $1+min(p^{e})$ MUB.

%%% \subsection{Ivanovic formulas for $n$ composite}\label{noirredcore}

The bound $1+\underset{i}{min} \{p_{i}^{e_i}\}$ can be easily reached for dimension $N=\prod_i p_{i}^{e_i}$.
It suffices to view $\mathbb{C}^N$ as $\otimes_{i}\mathbb{C}^{p_{i}^{e_i}}$.
As $\langle b_i\otimes c_k | b_j\otimes c_l\rangle=\langle b_i | b_j\rangle .\langle c_k |c_l\rangle$ we may conclude that a tensor product of two sets with $t$ MUB is a set of $t$  MUB in the product space.
Since there exist at least $1+\underset{i}{min} \{p_{i}^{e_i}\}$ MUB in each $\mathbb{C}^{p_{i}^{e_i}}$
we may construct by tensor product of these, a set of $1+\underset{i}{min} \{p_{i}^{e_i}\}$ MUB in the product space  $\mathbb{C}^N$.

%If $\{b_i\}_{i=1\ldots n}$ and $\{c_i\}_{i=1\ldots k}$ are orthonormal  bases
%respectively in $\mathbb{C}^n$ and $\mathbb{C}^k$,  then the product set
%$\{ b_i \otimes b_j\}$ is an orthonormal basis of $\mathbb{C}^{n.k}$.
%If we have a pair of mutually unbiased bases  $\{ B_1, B_2\}$ in $\mathbb{C}^n$ and a pair $\{ C_1, C_2\}$ in
%$\mathbb{C}^k$ then the bases $B_1\otimes C_1$ and $B_2\otimes C_2$ are mutually unbiased in
%$\mathbb{C}^{n.k}$ because
% $\langle b_i\otimes c_k | b_j\otimes c_l\rangle=\langle b_i | b_j\rangle .\langle c_k |c_l\rangle .$
%Since there exist at least $\underset{i}{min} \{p_{i}^{e_i}\}$ in each $\mathbb{C}^{p_{i}^{e_i}}$
%we may construct by tensor product of these, a set of $\underset{i}{min} \{p_{i}^{e_i}\}$ in the product %space  $\mathbb{C}^N$.

\subsection{Discussion on larger generalizations and conclusion}
In order to further generalize formula (\ref{generalize}) it could be tempting to allow the index set $S$
to be any subset of a finite ring. Unfortunately, this leads to a situation where any set of vectors could be described by such a formula.
To see this, let us recall that functions $f$ that preserve a decomposition of a ring $R$
are arbitrary functions on each component $R_i$. If we choose a ring $R$ that has no decomposition
(a field for instance) then such a $f$ is arbitrary on $R$. If $S$ is a subset $\{s_1,\ldots,s_N\}$ of $N$ elements, then
we may associate an arbitrary set of vectors $\{v_k\}_{1\leq k\leq N}$ in $\mathbb{C}^N$ to the couples in $S\times S$ by $(s_k,s_l)\to (v_k)_l.$

This may be extended (in many ways) to a two variables function  from $R\times R$ into $\mathbb{C}$ that preserves every decomposition of $R$ (since $R$ cannot be decomposed). One cannot expect to
reach algebraic conclusions that are valid for all $N\times N$ arrays with arbitrary complex entries $(v_k)_l$.\\

For these reasons, it is difficult to generalize formula (\ref{generalize}) much more.
% is a generalization that is large enough to indicate that,
It indicates that
for dimensions that are not prime powers, algebraic formulas providing complete sets of MUB should have a radically new structure. However, do these complete sets exist for any dimension ~?
Mathematicians are used to properties that behave differently for some particular dimensions
 but  such an answer is unsatisfactory from a physical point of view. W.K. Wooters has showed that the
absence of $N+1$ MUB for a dimension $N$ would be problematic for defining a discrete Wigner function in
systems having $N$ degrees of freedom (see \cite{wigner}). A negative answer to the MUB problem might
have other physical consequences and these could be used to guide mathematical investigations.

\section*{Acknowledgement}
I gratefully acknowledge Sofyan Iblisdir for introducing me to this problem and for
all our fruitful discussions.

\end{document}